\documentclass[10pt]{article}
\usepackage[OE]{express}

\newcommand{\p}{\partial}

\newcommand{\om}{\omega}
\newcommand{\nn}{\nonumber}
\newcommand{\ta}{\theta}

\begin{document}

\title{Multistability and coexisting  soliton combs in ring resonators: the Lugiato-Lefever approach}

\author{Y.V. Kartashov,\authormark{1,2,3} O. Alexander,\authormark{1} and D.V. Skryabin\authormark{1,4,*}}

\address{
\authormark{1}Department of Physics, University of Bath, Bath BA2 7AY, UK\\
\authormark{2}
ICFO-Institut de Ciencies Fotoniques, The Barcelona Institute of Science and Technology, 08860 Castelldefels (Barcelona), Spain\\
\authormark{3} 
Institute of Spectroscopy, Russian Academy of Sciences, Troitsk, Moscow Region, 142190, Russia\\ \authormark{4} Department of Nanophotonics and Metamaterials, ITMO University, St. Petersburg 197101, Russia}

\email{\authormark{*}d.v.skryabin@bath.ac.uk} 



\begin{abstract}
We are reporting that the Lugiato-Lefever equation describing the frequency comb
generation in ring resonators with the localized pump and loss terms also describes the simultaneous
nonlinear resonances leading to the multistability of nonlinear modes and coexisting solitons that are associated with the spectrally distinct frequency combs.
\end{abstract}



\section{Introduction}
Generation of frequency combs spanning an optical octave \cite{del_o} and the soliton formation \cite{her,yi,leo} are the  primary nonlinear effects in the microring  and in the passive fiber-loop ring resonators currently attracting  a growing attention. Both of these effects and both of these systems are closely related and are connected to a variety of applications, such as e.g., all-optical signal processing \cite{pfe} and high precision spectroscopy \cite{suh}. Solitons in the ring resonators circulate indefinitely providing  losses  are compensated by the continuous wave pumping. On the contrary, the solitons freely propagating in optical fibers and other waveguides are, and in their essence, quasi-solitons since the unbalanced losses and higher order effects lead to the energy radiation \cite{skr}.

A classic Lugiato-Lefever equation (LLE) \cite{lug} is widely used  to model the frequency combs and solitons in ring resonators, see, e.g., \cite{her,che}. If one sweeps the pump frequency  over the range of several resonances, then the LLE successfully describes the nonlinear tilt and the associated bistability effect only for a single resonance frequency, see, e.g., \cite{her,che,mil}. This is  an obvious shortcoming since doing such a sweep in an experiment produces a bistable response at every resonance. An alternative approach  is to apply the Ikeda map model, which was first used for the solitons in ring resonators a few decades ago \cite{mol}. The Ikeda approach was extended to the nonlinear fiber rings \cite{hae2} and more recently to the microring comb generation, where it predicted  several nonlinear resonances, multistability and {\em supersolitons} \cite{han_f}. The title of Ref. \cite{han_f}  explicitly says that these  effects are beyond what the LLE can describe. Disadvantages of the Ikeda approach is that the periodic boundary conditions are imposed in the evolution coordinate, which complicates mathematical methods and physical interpretation of even the simplest stability analysis of a homogeneous state \cite{hae2,han_f}, which is standard and routine in the LLE, see, e.g., \cite{her,mil}.  

The LLE typically utilizes the space independent pump term to describe frequency combs and associated solitons, see, e.g., \cite{her,che,mil,lug}, while the Ikeda approach intrinsically relies on the pump that is localized at one point \cite{mol,hae2,han_f}. The salient feature of the latter is that its expansion into the resonator modes has an infinite spectrum of the Dirac-delta function. Thus, in the LLE methods used so far the pump term has nonzero projection on only one mode, while the Ikeda method implies equal pump strength for the entire spectrum of the resonator modes. Clearly, both of these limits are mathematical idealizations, which have their regions of validity. In this work, we demonstrate how the LLE approach can be used to describe the multiple nonlinear resonances leading to the multistability effects and coexisting solitons with different widths and amplitudes that are associated with the spectrally distinct  frequency combs.
\begin{figure}
\centering
\includegraphics[width=0.9\textwidth]{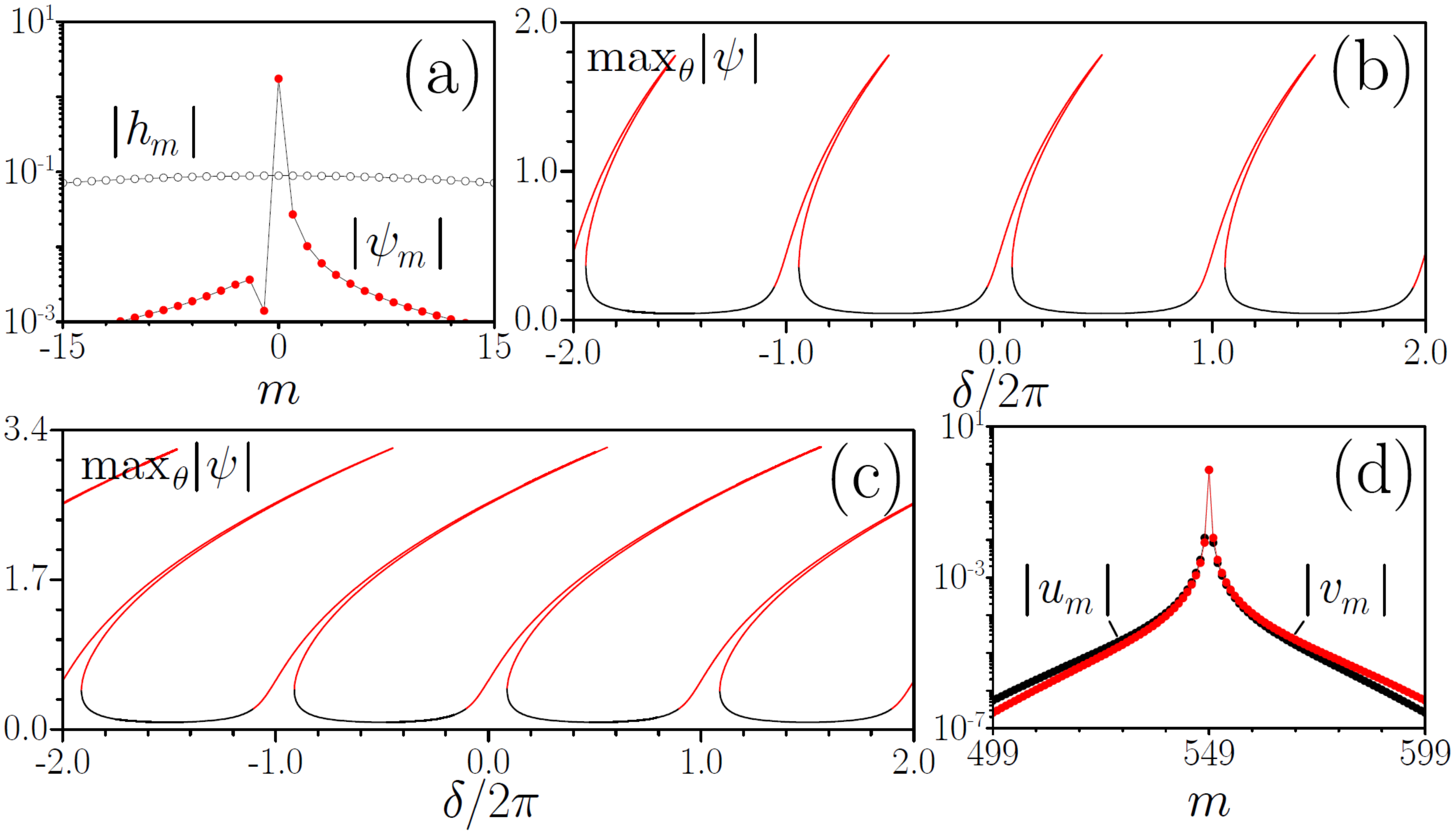}
\caption{(a) Spectra of the pump $H$ (empty circles) and of the nonlinear solution $\tilde\psi(\theta)=\sum_m\psi_me^{i\theta m}$ (red dots) corresponding to the upper branch of the $m=0$ resonance calculated for $h=5$, $\delta=3$, $\ta_0=\pi/50$, and $\gamma=0.05$, see Fig. 
\ref{f1}(c). (b,c) Multiple nonlinear resonances: max$_{\ta}|\tilde\psi(\ta)|$ vs the detuning parameter $\delta$. $h=5$ in (b) and $9$ in (c). (c) shows the nonlinear resonance tilts exceeding $2\pi$ and the associated multistability. (d) Spectra of the maximally unstable solutions of Eqs. (\ref{eq5}), 
$\{u(\ta),v(\ta)\}=\sum_m\{u_m,v_m\}e^{i\ta m}$, that drives the instability of the nonlinear state shown in (a). Note the log scales in (a) and (d).}
\label{f1}
\end{figure}
\section*{Lugiato-Lefever equation with the localised pump and loss}
The LLE model, well established in the context of the ring resonators \cite{her,yi,che,mil}, reads as
\begin{equation} i\p_T\Psi=(\omega_0-i\omega'\p_Z-\frac{1}{2}\omega''\p^2_Z)\Psi-i\kappa\Psi-T_r^{-1}|\Psi|^2\Psi-Pe^{-i\omega_p t}. \label{eq1}
\end{equation}
It describes the evolution of the dimensionless electric field amplitude $\psi$ in time $T$ and along the spatial coordinate  $Z$. All the parameters and variables in the above equation, apart from  $\Psi$ itself, are measured in physical units. The dimensionless amplitude $\Psi$  obeys an obvious and important periodicity condition $\Psi(T,Z-L/2)=\Psi(T,Z+L/2)$, where $L$ is the resonator length. Here $\kappa$ is the loss rate, $P$ and $\omega_p$ are the pump rate and the pump frequency, $T_r^{-1}|\Psi|^2$ is the nonlinear shift of the resonances. The coefficient before $|\Psi|^2$ has the dimension of the inverse time, but its value can be scaled to any convenient number, through the scaling of $P$ and $\Psi$ only. Our choice was to make it equal to the inverse round trip time,  $T_r=L/\omega'$. 

Disregarding the pump, loss and nonlinearity and assuming $\Psi=e^{im\beta Z-i\omega_mT}$, where $|m|=0,1,2,3,\dots$, one  finds
for the frequency of the m'th mode $\omega_m=\omega_0+\omega'm\beta+\omega''m^2\beta^2/2$. Here $\omega'$ is the group velocity and $\omega''$ is the dispersion coefficient at the frequency of the reference $m=0$ mode. $\omega''\ne 0$ makes the cavity spectrum slightly non-equidistant. In our case the dispersion is anomalous $\omega''>0$, see, e.g., Refs. \cite{her,mil} for the equations connecting the waveguide and the resonator dispersion coefficients.  The condition $\beta L=2\pi$ comes from the periodicity of $\Psi$. $\beta=2\pi/L$ has the meaning of the intermodal distance in the Fourier space reciprocal to the coordinate $z$, while $\omega'\beta$ is the free spectral range in the frequency space. In a typical setup, pump $P=P(Z)$ and losses $\kappa=\kappa(Z)$ are the functions of $Z$ that have a sharp maximum in the vicinity of a point where the resonator is coupled to the bus waveguide, which is used to  couple in the pump  and to couple out the signal. The effects of multistability and co-existence of different solitons studied below are introduced into the LLE model through the spatial inhomogeneity of the pump $P$  and are the primary subjects of this Express communication. 

In order to transform our system into a dimensionless form we assume  $Z=\theta L/(2\pi)$. Here $\theta$ is a new dimensionless coordinate, $\theta\in (-\pi,+\pi)$, which in the case of a ring shaped resonator is exactly the polar angle. Introducing $\tau=T/T_r$, $\delta=(\omega_0-\omega_p)T_r$, $\kappa T_r\equiv \Gamma(\theta)=\gamma (1+e^{-\theta^2/\theta_0^2})$, $P T_r\equiv H(\theta)=he^{-\theta^2/\theta_0^2}$ and $\Psi(T,Z)=\psi(\tau,\theta) e^{-i\om_pT}$ we transform Eq. (\ref{eq1}) into the convenient dimensionless form 
\begin{equation} i\p_{\tau}\psi=(\delta-i2\pi\p_{\theta}-d\p^2_{\theta})\psi-i\Gamma(\theta)\psi-|\psi|^2\psi-H(\theta),\label{eq2}\end{equation} 
$\psi(\theta=-\pi,\tau)= \psi(\theta=+\pi,\tau)$.  
Here $h$ and $\gamma$ are the maximal pump and loss values at $\theta=0$, $\theta_0\ll 2\pi$ is the angular width of the pump spot. The $\theta$-dependent separation between the straight pump waveguide and the ring shaped resonator waveguide is well approximated by $l+L\theta^2/(2\pi)$, where $l$ is the minimal separation at $\ta=0$. The coupling strength between the two waveguides is known to decay exponentially with the separation, which gives the Gaussian profile for $H(\theta)$. Note, that if one transforms Eq. (\ref{eq2}) into a reference frame making one round trip per unit time,  then the 
$\p_{\theta}$-term cancels out, but the pump and loss terms become periodic in time, providing their spatial inhomogeneity was accounted for: 
\begin{equation} i\p_{\tau}\psi=\delta\psi-d\p^2_x\psi-i\Gamma(x+2\pi \tau)\psi-|\psi|^2\psi-H(x+2\pi \tau),~x\equiv\theta-2\pi \tau. \label{eq3} \end{equation}
The dimensionless resonator dispersion parameter $d$ 
is the ratio between the round trip time and the dispersion time of a pulse having the width of the resonator length: $d=2\pi^2\frac{L/\omega'}{L^2/\omega''}$.  The soliton width scales as $(\Delta/d)^{1/2}$, which for  $d=10^{-5}$ allows to fit $\sim 1000$ solitons over the cavity length. Here $\Delta$ is the relative detuning from a chosen resonance, $0<\Delta\lesssim 2\pi$. Our choice of the relatively high resonator dispersion is convenient to reduce computational requirements, since the higher dispersion increases the width of the solitons relative to the resonator length, but it does not alter general applicability of our results to the resonators with a broad range of  $\omega''$ values, since we can use $L$ as a fitting parameter. The resonator finesse $F$ can be defined as the ratio between the free spectral range (FSR) and the cavity loss rate: $F=\frac{\omega'\beta}{\mathrm{min}~\kappa}=\frac{2\pi}{\gamma}$. We choose $\gamma=0.05$ in what follows, which gives the relatively low $F\simeq 125$. Relatively low $F$ achieved primarily through the small FSR are critical for observing multistability  and coexisting comb solitons discussed below. While our aim here is  to introduce a proof of principle theoretical results, the explicit scaling we presented above allows to relate our approach to a number of geometries and material choices. A practical implementation of the regimes considered here is likely to be possible, e.g., in the fiber loop resonators \cite{and} and in the  low finesse relatively long microrings \cite{bog}. 
\begin{figure}
\centering
\includegraphics[width=0.9\textwidth]{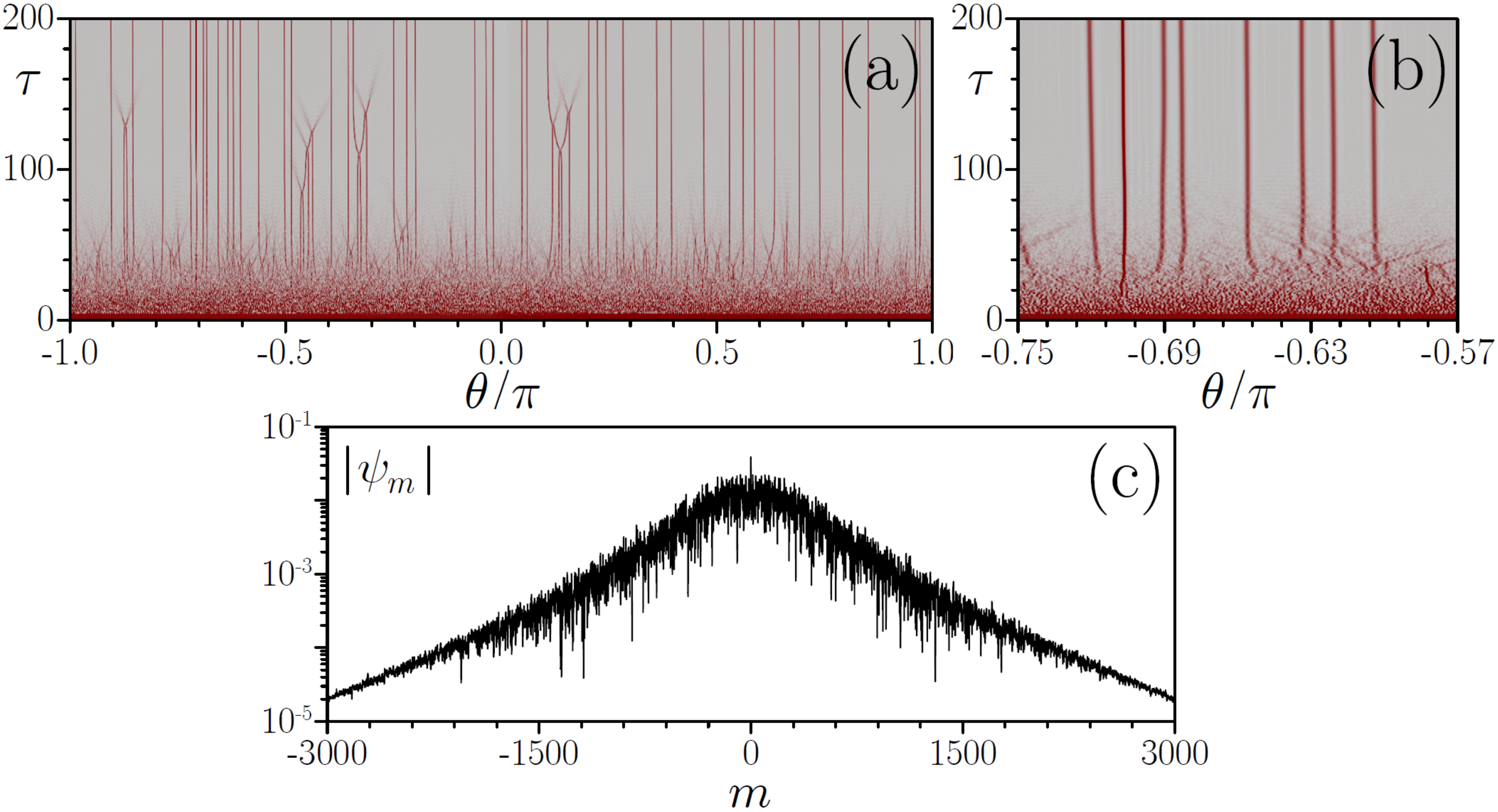}
\caption{(a) Spatiotemporal dynamics of comb formation observed in numerical modeling of Eq. (\ref{eq2}) initialized by the highest amplitude solution in the multistability regime shown in Fig. \ref{f1}(c) at $\delta=8.4$. (b) shows a zoom of the $\theta\in [-0.75\pi,-0.57\pi]$ interval. (c) shows the associated comb spectrum at $\tau=200$.}
\label{f2}
\end{figure}
\begin{figure}
\centering
\includegraphics[width=0.9\textwidth]{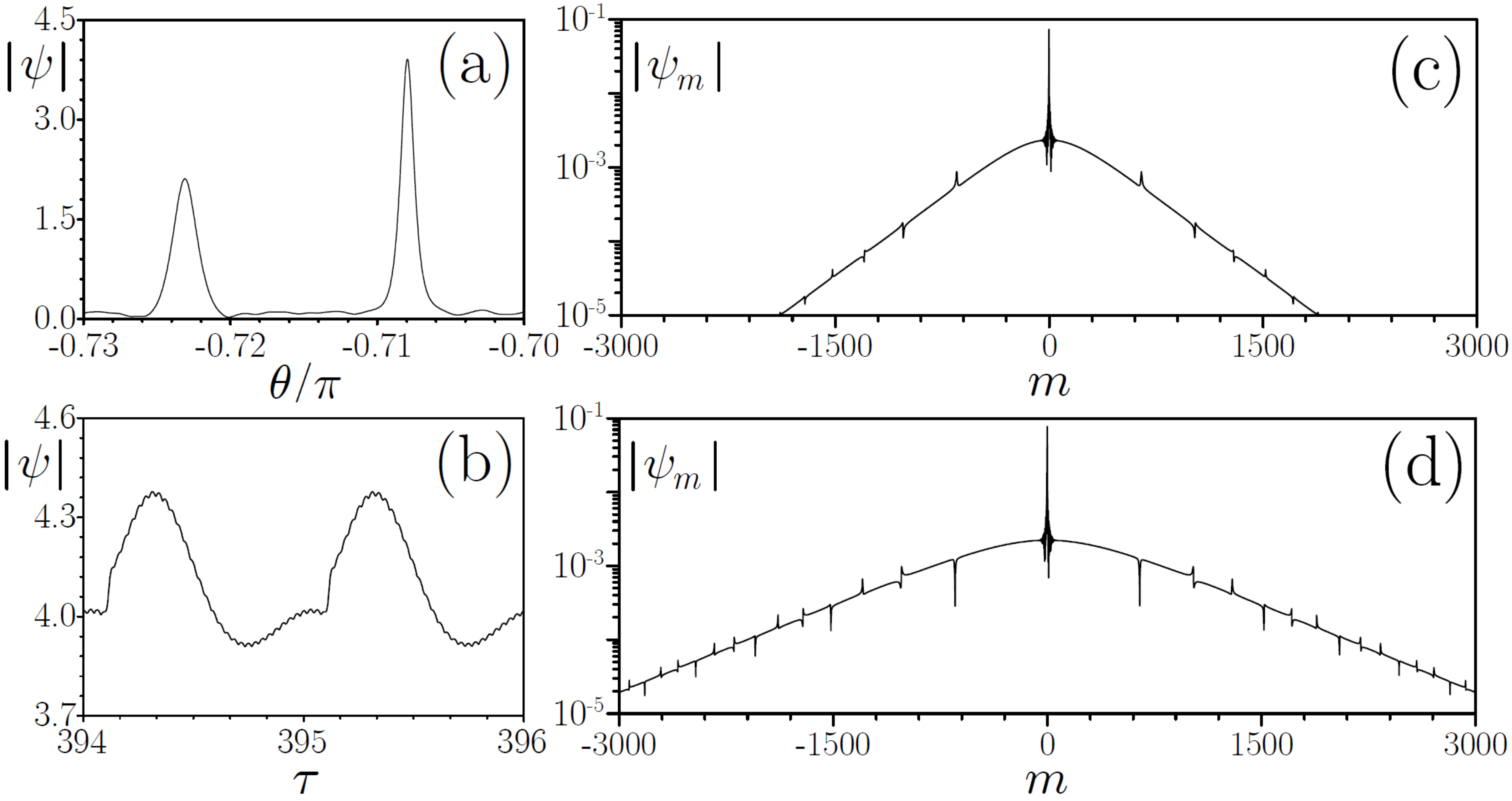}
\caption{(a) A zoomed interval of $\theta$ from Fig. \ref{f2}(a) showing the coexisting comb solitons with the different widths and amplitudes. (b) 
Breathing of the amplitude of the larger soliton from (a) as it makes round trips around the resonator.
(c,d) The comb spectra of the isolated low (c) and high (d) amplitude solitons from (a).}
\label{f3}
\end{figure}
\section{Linear and nonlinear modes, multistability and instabilities}
We first consider stationary, $\p_{\tau}=0$, nonlinear modes of Eq. (\ref{eq2}). We expand $\psi(\theta,\tau)=\tilde\psi(\theta)$, $\Gamma(\theta)$ and $H(\theta)$ into Fourier series $\{\tilde\psi(\theta),\Gamma(\theta),H(\theta)\}=\sum_{m=-\infty}^{\infty}\{\psi_m,\gamma_m,h_m\}e^{im\theta}$ and obtain the following algebraic system:
\begin{equation} (\delta+2\pi m+dm^2)\psi_m=i\sum_n\gamma_n\psi_{m-n}+\sum_{kn}\psi_k\psi_n\psi^*_{m-k+n}+h_m.
\label{eq4}\end{equation}
When losses are uniform and nonlinearity is absent the single mode solutions  are given by 
$\psi_m=h_m/(\delta +2\pi m+dm^2-i\gamma)$. Plotting $\psi_m$ as a function of $\delta$ one finds a set resonances separated roughly by $2\pi$, with a small offset due to $d\ne 0$. Since the amplitude of the  $m$-th resonance is proportional to  $h_m$, the resonance peaks are located under an envelope with the spectral width 
$\sim\theta_0^{-2}$. If the pump is uniform, then only the $m=0$ resonance has a nonzero amplitude, simply because the pump projection  on all other modes is zero. The positive Kerr nonlinearity tilts this resonance towards large $\delta$ introducing a bistability loop. 

Taking as an example $\theta_0=\pi/50$ (variation of $\theta_0$ 
within the realistic interval of few degrees produces quantitatively similar results) and accounting for $2000$ modes, we solved Eqs.  (\ref{eq4}) numerically using Newton method. 
Figs. \ref{f1}(b) and \ref{f1}(c) show the dependencies of the amplitude of the nonlinear mode $\tilde\psi(\theta)=\sum_m\psi_me^{im\theta}$ defined as max$_{\ta}|\tilde\psi(\ta)|$ on the detuning parameter $\delta$. One can see the multiple   tilted resonances, that start overlapping for the sufficiently strong pump and form what we call the multistability or tristability effect, see Fig. \ref{f1}(c). Spectra of the pump $H$ and  of the nonlinear solution $\tilde\psi$ are shown in Fig. \ref{f1}(a). $\tilde\psi$ is generally asymmetric with respect to the $\theta=0$  point in the real space and with respect to $m=0$ in the Fourier space due to the simultaneous presence of the $\p_{\theta}-$term   in Eq. (\ref{eq1}) and of the non-uniformities of $H$ and $\Gamma$. At the lowest  branches of the nonlinear resonance curves, this asymmetry is almost negligible, but it becomes noticeable for the high amplitude solutions, see Fig. \ref{f1}(a).

By taking the real-space Eq. (\ref{eq2}) we have linearised it around $\tilde\psi$ using the substitution $\psi(\theta,\tau)=\tilde\psi(\theta)+u(\theta)e^{\lambda\tau}+v^*(\theta)e^{\lambda^*\tau}$ with $|\tilde\psi|\gg |u|,|v|$:
\begin{eqnarray}
&& \lambda u=-i\delta u-2\pi\p_{\ta}u+id\p^2_{\ta}u-\Gamma u+2i|\tilde\psi|^2u+i\tilde\psi^2v\label{eq5}\\ \nn &&
\lambda v=i\delta v-2\pi\p_{\ta}v-id\p^2_{\ta}v-\Gamma v-2i|\tilde\psi|^2v-i\tilde\psi^{*2}u
\end{eqnarray}
The intervals of the instabilities calculated numerically from Eqs. (\ref{eq5}) are shown in Figs. \ref{f1}(b) and \ref{f1}(c) with the red lines. Similarly to the case of the uniform pump and losses, the upper branches of every resonance are unstable \cite{lug}. For every $\delta$ there is a finite set of the unstable $u,v$ pairs \cite{lug}. The spatial profiles of these pairs are dominated by the linear modes $e^{im\theta}$ with the modal numbers $m$ spanning a finite interval. Representative spectra of $u(\theta)=\sum_{m}u_me^{i\theta m}$ and $v(\theta)=\sum_{m}v_me^{i\theta m}$ corresponding to the maximal instability growth rate found for a given $\delta$ are shown in Fig. \ref{f1}(d). The instability in this particular example is dominated by the growth of the  $m=549$th mode. 
\section{Coexisting solitons and frequency combs}
It is well known by now that the instability of the upper branch of the nonlinear resonance leads to the generation of the frequency combs through the cascaded four-wave mixing process, see, e.g., \cite{che}. As $\delta$ approaches its value corresponding to the rightmost point of the tilted resonance the generally complex space-time comb signals are replaced by sets of  spontaneously generated  solitons, see, e.g., \cite{her,mil}. We  observed this dynamics in our system too. However, when the multistability is present, then we have an option of starting a simulation, for a given $\delta$, from either of the two coexisting upper branch solutions. Note, that the stable lowest amplitude solution can also coexist with the two unstable upper branches. The former provides a stable background for the soliton pulses with the peak amplitudes close to either the highest or the intermediate nonlinear resonances. 

When we performed simulations of the comb generation using Eqs. (\ref{eq2}) with $2^{16}$ modes and initializing them with the highest amplitude nonlinear mode found from the time independent Eqs. (\ref{eq4}), we have typically observed the dynamics leading to the comb generation through the formation of multiple solitons, see Fig. \ref{f2}(a). The  associated frequency comb  is shown in Fig. \ref{f2}(b). A close examination reveals that the solitons constituting this comb can have  two different amplitudes and widths.
Extensive numerical simulations show that output field distributions always contain a considerable number of the low-amplitude solitons and a few of the high-amplitude ones. Fig. \ref{f3}(a) 
shows a zoomed interval of $\ta$'s from Fig. \ref{f2}(a) capturing the two different solitons stably propagating one next to the other. The taller one is also narrower since it corresponds to the larger effective detuning from the associated linear resonance. Isolating either of the two solitons from the rest of the signal, we have observed that both of them persist in the resonator practically indefinitely. 
The comb associated with the high amplitude soliton is much broader, cf. Figs. \ref{f3}(c) and \ref{f3}(d). If one  compares spectra of the individual solitons  with the spectrum of the whole signal in Fig. \ref{f2}(b), one can clearly see the two different scales in tails of the latter. Thus  the central part of the spontaneously generated comb is associated with the low-amplitude solitons, while the outer wings are determined by the high-amplitude narrow-width ones.

The comb solitons reported here, both inside and outside the multistability regime,  reside on top of the oscillating in space and time background wave and demonstrate pronounced breathing. The breathing soliton amplitude is shown in Fig. \ref{f3}(b). The oscillation period matches the resonator round trip time and hence is associated with the localizations of the pump and loss terms. However, the maxima and minima of these amplitude oscillations do not generally coincide with the moments when the soliton overlaps with the maximum of $H$. Also, the amplitude of the oscillations varies slightly even for the solitons of the same type, but located at different points. This temporal dynamics is associated with  the modulation of the comb soliton spectra in Figs. 3(c) and 3(d), which is similar to the spectral modulation reported in Ref. \cite{np1}.

\section{Conclusions} We demonstrated how the LLE model describing the frequency comb generation can be generalized to predict  multiple simultaneous nonlinear resonances, multistability and coexistence of the small and high amplitude solitons and of the frequency combs with different spectral widths.

\section*{Funding} The Leverhulme Trust (RPG-2015-456); H2020 (691011, Soliring); ITMO University (Grant 074-U01); RFBR (17-02-00081); RSCF (17-12-01413).

\begin{thebibliography}{99}

\bibitem{del_o}
P. Del'Haye, T. Herr, E. Gavartin, M. L. Gorodetsky, R. Holzwarth, and T. J. Kippenberg,  "Octave spanning tunable frequency comb from a microresonator," Phys. Rev. Lett. {\bf 107}, 063901 (2011).

\bibitem{her}
T. Herr, V. Brasch, J. D. Jost, C. Y. Wang, N. M. Kondratiev, M. L. Gorodetsky, and T. J. Kippenberg, "Temporal solitons in optical microresonators," Nature Photon. {\bf 8}, 145 (2014).



\bibitem{yi}
V. Brasch, M. Geiselmann, T. Herr, G. Lihachev, M. H. P.  Pfeiffer, M. L.  Gorodetsky,  and T. J.  Kippenberg,  
"Photonic chip-based optical frequency comb using soliton Cherenkov radiation," Science {\bf 351}  357-360 (2016).

\bibitem{leo} F. Leo, S. Coen, P. Kockaert, S. P. Gorza, P. Emplit, and M. Haelterman, "Temporal cavity solitons in one-dimensional Kerr media as bits in an all-optical buffer," Nature Photon. {\bf 4}, 471 (2010).

\bibitem{pfe}
J. Pfeifle,  V. Brasch, M. Lauermann, Y. Yu, D. Wegner, T. Herr, K. Hartinger, P.  Schindler, J.S. Li, D. Hillerkuss, R. Schmogrow, C. Weimann, R. Holzwarth, W. Freude, J. Leuthold, T.J. Kippenberg, and C. Koos,  "Coherent terabit communications with microresonator Kerr frequency combs," Nat. Photon. {\bf  8},  375-380 (2014).

\bibitem{suh}
M. G. Suh, Q. F. Yang,  K. Y. Yang,   X. Yi,  and K. J. Vahala, "Microresonator soliton dual-comb spectroscopy," Science {\bf 354},   600-603 (2016).

\bibitem{skr}
D. V. Skryabin and A. V. Gorbach, "Looking at a soliton through the prism of optical supercontinuum," Rev. Mod. Phys. {\bf 82}, 1287-1299 (2010).

\bibitem{lug} L. A. Lugiato and R. Lefever, "Spatial dissipative structures in passive optical systems," Phys. Rev. Lett. {\bf 58}, 2209 (1987).

\bibitem{che} 
Y. K. Chembo and N. Yu, "Modal expansion approach to optical-frequency-comb generation with monolithic whispering-gallery-mode resonators," Phys. Rev. A {\bf 82}, 033801 (2010);
Y. K. Chembo and C. R. Menyuk, "Spatiotemporal Lugiato-Lefever formalism for Kerr-comb generation in whispering-gallery-mode resonators," Phys. Rev. A {\bf 87}, 053852 (2013).

\bibitem{mil}	
C. Milian, A. V.  Gorbach, M. Taki, A. V. Yulin,  and D. V. Skryabin, "Solitons and frequency combs in silica microring resonators: Interplay of the Ramanand higher-order dispersion effect," Phys. Rev.  A {\bf 92}, 033851 (2015).

\bibitem{mol} D. W. McLaughlin, J. V. Moloney, and A. C. Newell, "Solitary waves as fixed points of ininetely dimensional maps in an optical bistable ring cavity," Phys. Rev. Lett. {\bf 51}, 75-78  (1983).


\bibitem{hae2} S. Coen and M. Haelterman, "Modulational instability induced by cavity boundary conditions in a normally dispersive optical fiber," Phys. Rev. Lett. {\bf 79}, 4139 (1997).


\bibitem{han_f}	T. Hansson and S. Wabnitz, "Frequency comb generation beyond the Lugiato-Lefever equation: multi-stability and super cavity solitons," J. Opt. Soc. Am. B {\bf 32}, 1259 (2015).

\bibitem{and}	M. Anderson, Y. Wang, F. Leo, M. Erkintalo, S. Coen, and S. G. Murdoch, "Coexistence of distinct cavity soliton states in a tristable passive Kerr resonator," 
in Photonics and Fiber Technology 2016 (ACOFT, BGPP, NP), OSA Technical Digest (online) (Optical Society of America, 2016), paper NW3B.1.


\bibitem{bog}
W. Bogaerts, P. De Heyn, T. Van Vaerenbergh, K. DeVos, S. K.
Selvaraja, T. Claes, P. Dumon, P. Bienstman, D. Van Thourhout, and R. Baets, "Silicon microring resonators," Laser Photon. Rev. {\bf 6}, 47-73 (2012).

\bibitem{np1}
J. K. Jang, M. Erkintalo, S. G. Murdoch, and S. Coen, "Ultraweak long-range interactions of solitons observed over astronomical distances," Nat. Photon. {\bf 7}, 657-663 (2013).

\end{thebibliography}
\end{document}